\documentstyle[aps,amssymb]{revtex}
\begin{document}
\title{
On extension of minimality principle\\
in supersymmetric electrodynamics
}
\author{
V.V. Tugai
\footnote{E-mail address:  tugai@sptca.kharkov.ua}
}
\address{
Scientific Physicotechnological Center, 310145 Kharkov, Ukraine
}
\author{
A.A. Zheltukhin
\footnote{E-mail address: kfti@rocket.kharkov.ua}
}
\address{
Kharkov Physicotechnical Institute, 310108 Kharkov, Ukraine
}
\maketitle

\quad

Deep geometric ideas of Eli Cartan \cite{Cartan1} inspired
Dmitrij Vasiljevitch Volkov in his investigations of
the phenomenological lagrangians method \cite{Volkov10}
and in his work at the construction of
supersymmetry and supergravity \cite{pionSUSY,SUSY_SGRA_10}.
The Cartan idea on the extension of the connection conception and the
introduction of torsion was applied by
Dmitrij Vasiljevitch Volkov in the formulation of minimality
principle for the interactions of Goldstone particles with other fields
\cite{SUSY_101}. His profound intuition at once allowed to assume
that the Pauli matricies $\sigma ^{\mu }_{\alpha \dot\alpha }$
play a fundamental role of the torsion tensor components in
$z^M = (x^{\mu }, \theta ^{\alpha }_{i},
\bar {\theta }_{\dot \alpha  i})$ superspace.

The Cartan's geometry
naturally emerged in early papers by Scherk and Schwarz \cite{ShSch1}
on the string theory of gravitation where the strength
$H_{\mu \nu}{}^{\rho }$ of the antisymmetric Kalb-Ramond field
plays the role of torsion \cite{Kalb}. These and some other
impressive results show ties between superspace torsion and spin.

Here we wish to single out a possibility of an extension of the
minimality principle for electromagnetic interactions of charged and
neutral particles having spin $1/2$.
This possibility is also based on the Cartan's idea of an extension
of the connection conception.

\section{}

An extention of the minimality principle can be achieved by the
addition of new terms \cite{TZFR2,TZPL3}
to the standard gauge covariant and supesymmetric
derivatives ${\cal D}_M$ of supersymmetric electrodynamics \cite{WESS}
\begin{equation}
{\cal D}_M = D_M + e A_M \quad \to \quad
\nabla _M = {\cal D}_M + i\mu \tilde W_M \equiv
D_M + e_{(q)}{\Bbb A}^{(q)}_M,
\label{to1}
\end{equation}
where the additional superfield
$\tilde W_M(z)$ is an invariant of the gauge group $U(1)$,
$e_{(q)}{\Bbb A}^{(q)}_M\equiv eA_M +i\mu \tilde W _M$
is a new generalized connection, and the constant $\mu $ with
the dimension of length ($\hbar = c =1$) has the physical meaning
of anomalous magnetic moment (AMM).

The algebra of the doubly lengthened covariant derivatives
$\nabla _M = (\nabla
_{\mu },$$\nabla _{\alpha }{}^i ,$$ \bar \nabla ^{\dot \alpha i})$
differs from the algebra of the standard $N$-extended
supersymmetric derivatives \cite{West}
\begin{eqnarray}
&& \left[ D_M , D_N \right\} = T_{MN} {}^L D_L,\cr
&&\cr
&& T_{MN} {}^L =-2i \sigma _{\alpha \dot\beta }^{\mu }, \qquad
\hbox{for $N=1$ SUSY} \cr
&& T_{MN} {}^L =2i \sigma _{\alpha \dot\beta }^{\mu } \delta _i^j, \qquad
\hbox{for $N>1$ SUSY}
\label{OrdSUSYDer}
\end{eqnarray}
by the presence of the electromagnetic superfield strength
${\Bbb F} ^{(q)}_{MN}$
\begin{equation}
\left[ \nabla _M , \nabla _N \right\} = T_{MN} {}^L \nabla _L
+ e_{(q)} {\Bbb F} ^{(q)}_{MN}.
\label{TwoLongSUSY}
\end{equation}
which takes into account AMM of superparticle and is defined as
\begin{eqnarray}
&&e_{(q)}{\Bbb F}^{(q)}_{MN} \equiv e F_{MN}+i\mu F^{(\mu )}_{MN}
,\cr
&&F_{MN} \equiv D_M A_N - (-1)^{MN} D_N A_M - T_{MN}{}^{R} A_R, \cr
&&F^{(\mu )}_{MN} \equiv D_M \tilde W_N - (-1)^{MN} D_N \tilde W_M -
T_{MN}{}^{R} \tilde W_R.
\label{FmnFull}
\end{eqnarray}

Conservation of supersymmetry and $U(1)$ gauge symmetry
demands that the superfield $\tilde W_M (z)$ should be an invariant of
these symmetries. Let us look for these invariants in
the set of functions which are a linear combination of the
strength components $F_{MN} (z)$ with constant coefficients.
The dimensionality reasons, i.e.
$[\tilde W_{\mu }]=L^{-2}$, $[\tilde W_{\alpha }^{i}]=L^{-3/2}$,
together with the constraints imposed on $F_{MN}$ \cite{WESS,West}
sharply restrict the form of the invariant
$\tilde W_M$.
As a result, the desired spinorial components of
$\tilde W_M$  are taken in the form of the product of
$F_{\mu \alpha }{}^i$  by the torsion of the flat superspace
$\tilde W_{\dot\alpha i} \sim F_{\mu }{}^{\alpha}{}_j
\sigma ^{\mu }_{\alpha \dot\alpha }\delta ^j_i$.
In particular, for $N=2$ case the expression
$\tilde W_{\mu } \sim \partial _\mu F_{\alpha }{}^{i}{}_{\beta }{}^j
\varepsilon ^{\alpha \beta } \varepsilon _{ij}$
may be taken for the vector component
of $\tilde W_{\mu }$.
Due to the symmetry of $F_{\alpha \beta }$ this expression
equals zero for $N=1$ case.
The desired representation for the superfield $\tilde W_M$
in $N=1$ case has the following form \cite{TZFR2,TZPL3}
\begin{equation}
\tilde W_{M} =W_{M} \equiv
{i\over 4} (0,-\sigma _{\mu\alpha{\dot\alpha}}F^{\mu{\dot\alpha}},
{\tilde\sigma}^{\mu{\dot\alpha}\alpha}F_{\mu\alpha}).
\label{WForDerivLong}
\end{equation}

The uniqueness of the representation (\ref{WForDerivLong})
arises when taking account of the standard $F_{MN}$ constraints
\cite{WESS}
\begin{eqnarray}
&&F_{\alpha \beta}=F_{{\dot \alpha}{\dot \beta}}=0,
\label{ConstrWessI}
\\
&&F_{\alpha {\dot \beta}} =0.
\label{ConstrWessII}
\end{eqnarray}
and the Bianchi identities
\begin{equation}
{\Bbb C}\hbox{ycl}_{MNR} (-1)^{(MNR)}
\left( D _M  F_{NR} - T_{MN}{}^L A _{R}\right) =0
\label{BiaSupInd}
\end{equation}
equivalent to the Jacobi identities for the covariant derivatives
${\cal D}_M$ of supersymmetric electrodynamics (SUSY ED)
\begin{equation}
{\Bbb C}\hbox{ycl}_{MNR} (-1)^{MNR}
\left[ {\cal D} _M ,\left[ {\cal D} _N ,
{\cal D} _R \right\} \right\} =0.
\label{JakInd}
\end{equation}

Eqs. (\ref{ConstrWessI},\ref{ConstrWessII}) require that the superfield
$F_{MN}$ of $N=1$ SUSY ED should be expressed in terms of the spinor
chiral superfields $W_{\alpha }$ and $\bar W _{\dot \alpha }$
\cite{WESS}
\begin{equation}
W^{\alpha }=
{i\over 4}F_{\mu\dot\alpha }\tilde\sigma ^{\mu \,\dot\alpha \alpha },
\qquad
\bar W^{\dot\alpha }= (W^{\alpha })^*.
\label{WfromFMN}
\end{equation}
Therefore $\tilde W_{\mu} $ must be constructed of $W_{\alpha }$
and $\bar W _{\dot \alpha }$.
Using these superfields one cannot construct the vector
superfield $\tilde W_{\mu} $ having the dimension
$L^{-2}$. As a result, the desired extension of the minimality
principle for $N=1$ SUSY ED is defined by the superfield
$\tilde W_{M} $ (\ref{WfromFMN}) having the form
\begin{equation}
\tilde W_M(z)=(0,W_{\alpha },\bar W ^{\dot\alpha })
\label{TmWm}
\end{equation}
with $W_{\alpha }$ and $\bar W _{\dot \alpha }$ restricted by the
chirality and reality conditions \cite{WESS}
\begin{eqnarray}
D_{\alpha} {\bar W}_{\dot \alpha}={\bar D}_{\dot \alpha}W_{\alpha}=
D^{\alpha } W_{\alpha }- {\bar D}_{\dot\alpha }{\bar W}^{\dot\alpha }=0.
\label{ConstrForWI_II}
\end{eqnarray}
The proof of the uniqueness is completed by the analysis of the
Jacobi identities
\begin{equation}
{\Bbb C}\hbox{ycl}_{MNR} (-1)^{(MNR)}
\left[ \nabla _M ,\left[ \nabla _N ,
\nabla _R \right\} \right\} =0 ,
\label{JakIndII}
\end{equation}
which take the form
\begin{eqnarray}
&&e{\Bbb C}\hbox{ycl}_{MNR}(-1)^{(MNR)}
\left( D _M  F_{NR} - T_{MN}{}^L F_{LR}\right) +
i\mu {\Bbb C}\hbox{ycl}_{MNR} (-1)^{(MNR)} \left( D _M  F^{(\mu )}_{NR}
- T_{MN}{}^L F^({\mu })_{LR}\right) =0
\label{BiaSupIndII}
\end{eqnarray}
after the substitution of Eqs.(\ref{to1}), (\ref{TwoLongSUSY}) and
(\ref{FmnFull}) into Eq. (\ref{JakIndII}).
The first and the second
term in Eq.(\ref{BiaSupIndII}) equal zero by
construction and due to sufficient arbitrariness
in the definition of
$W_{\alpha }$ and $\bar W _{\dot \alpha }$, respectively

Therefore the extended derivatives (\ref{to1}, \ref{TmWm})
satisfy the Bianchi identities and give a solution of the
desired problem of selfconsistent extension of the
minimality principle for $N=1$ SUSY ED particles with
spin $1/2$.

The superalgebra (\ref{TwoLongSUSY}) of the generalized
derivatives $\nabla _M$ takes the form
\begin{eqnarray}
[\nabla _\mu , \nabla _\nu ] &=&
- {e\over 2} (\bar D \tilde\sigma _{\mu \nu }\bar W -
D \sigma _{\mu\nu} W ),\cr
[\nabla _\mu , \nabla _\alpha  ] &=&
i e W^{\beta }\sigma _{\mu\beta \dot\alpha}
+i\mu \partial _\mu W _\alpha  ,\cr
\{\nabla _\alpha  , \nabla _\beta \} &=& i\mu
(D_\alpha  W _\beta + D_\beta W_\alpha )  ,\cr
\{\nabla _\alpha  , \bar\nabla _{\dot\beta} \} &=&
-2i \sigma ^\mu _{\alpha \dot\beta } \nabla _\mu+
i\mu (D_\alpha  \bar W _{\dot\beta} + \bar D_{\dot\beta} W_\alpha ),
\label{TwoLongDerAlg}
\end{eqnarray}
and we observe the noncommutativity effect for the generalized spinor
derivatives $\nabla _\alpha $.

Due to the considered extension procedure, the action for
charged or neutral superparticle in an external electromagnetic
field gets an additional term which describes the electromagnetic
interaction of superparticle taking into account its AMM
\begin{eqnarray}
&&
S_{int}=i \int dz^M e_{(q)} {\Bbb A}_M ^{(q)} = S^{(e)} +S^{(\mu)},
\cr
&&
S^{(e)} = ie \int d z^M A_M,
\cr
&&
S^{(\mu )} = -\mu \int dz^M \tilde W_M
=
- \mu \int d \theta ^{\alpha } W_{\alpha }
- \mu \int d \bar \theta _{\dot\alpha} \bar W ^{\dot\alpha}
.
\label{toL}
\end{eqnarray}
The component analysis of the action (\ref{toL})
carried out in \cite{TZFR2,TZPL3} shows that $S^{(\mu )}$
term in Eq.(\ref{toL}) generates the nonminimal Pauli term
with the constant $\mu $ which may be interpreted as the AMM of Dirac
particle measured in values of Bohr magneton.

In the next section we consider a generalization of the extended
minimality principle for the case $N=2$.

\section{}

In the standard $N=2$ SUSY ED \cite{Lusanna,Sohnius}
the strength components $F_{MN}$ obey the constraints
\begin{eqnarray}
&&F_{(\alpha }{}^{i}{}_{\beta )}{}^{j}=
F_{(\dot\alpha |i|\dot\beta )j}=0,
\label{ConstrWessN2I}
\\
&&F_{\alpha }{}^{i}{}_{\dot\beta j}=0.
\label{ConstrWessN2II}
\end{eqnarray}
Together with the Bianchi identities, these constraints permit to express
$F_{MN}$ in terms of the scalar chiral superfield
$W(z)$ and $\bar W(z)$ having the dimension $L^{-1}$ and restricted by
the conditions
\begin{equation}
\bar D _{\dot\alpha i}W=D_{\alpha }{}^{i}\bar W =0.
\label{ConWN2I}
\end{equation}
The explicit form of the superfield $F_{MN}$ components is
given by the following expressions
\begin{eqnarray}
&&F_{\alpha }{}^i {}_{\beta }{}^j = -\varepsilon _{\alpha \beta }
\varepsilon ^{ij}\bar W, \quad
F_{\dot\alpha i \dot\beta j} = \varepsilon _{\dot\alpha \dot\beta }
\varepsilon _{ij}W ,
\cr
&&F_{\mu \alpha }{}^{i} = {i\over 4}(\sigma _{\mu }\bar D ^i)
_{\alpha }\bar W, \
F_{\mu \dot\alpha i} = -{i\over 4}(D_i \sigma _{\mu })_{\dot\alpha}W,
\cr
&&F_{\mu \nu }=
{1\over 16}(\bar D _i \tilde \sigma _{\mu\nu  }\bar D^i)\bar W +
{1\over 16}(D^i \sigma _{\mu\nu  } D_i) W .
\label{FfromWN2}
\end{eqnarray}
In order to construct an extended connection for the $N=2$ SUSY ED,
we must again use the dimensionality reasones and look for the required
extension in terms of a linear combination of $W$, $\bar W$ and
their covariant derivatives. Then the required representation for the
additional superfield $T_M$ (which produces additional term to the
connection $A_M$) may be written in the form
\begin{equation}
\tilde W_M (z) = T_M
= (\partial _\mu (b W + \bar b  \bar W),\,
a D_{\alpha }{}^{i}W, \,
- \bar a \bar D^{\dot\alpha i}\bar W ).
\label{TMforN2}
\end{equation}
In a way analogous to that used in $N=1$ case
\cite{TZFR2,TZPL3} it is convenient to introduce the two-component
``charge''
$e_{(q)}=(e,i\mu )$  and the superconnection
${\Bbb A}_M^{(q)}=\left( {A_M\atop
T_M}\right) $ in the extended ``charged'' space $e_{(q)}$.
Then the doubly lengthened covariant derivative
(\ref{TwoLongSUSY}) can be more compactly presented as
\begin{eqnarray}
\nabla _{M} = D_{M} +e_{(q)} {\Bbb A}_M^{(q)}.
\label{TwoLongDTwoC}
\end{eqnarray}

Then the $U(1)$ gauge invariant superfield action for
charged or neutral particles with AMM in an external superfield of
the $N=2$ Maxwell supermultiplet takes the form
\begin{eqnarray}
\hbox{S}_{\hbox{int}} &=&
\hbox{S}_{\hbox{int}}^{(e)} + \hbox{S}_{\hbox{int}}^{(\mu )} =
i e_{(q)} \int \limits_{\Gamma ^*} \omega ^M (dz) {\Bbb A}_M^{(q)}
= i \int\limits _{\Gamma ^*} (e \omega ^M A_M + i\mu \omega ^M T_M),
\label{SN2int1}
\end{eqnarray}
where $N=2$ Cartan's differential forms
$\omega ^{M}(dz)=(\omega ^{\mu } ,\omega ^{\alpha }{}_i ,
\omega ^{\dot\alpha i})$
in $N=2$ superspace $z_M$ are given by \cite{pionSUSY,West}:
\begin{eqnarray}
&&\omega ^{\mu  }  = dx^{\mu } -i (d\theta _i
\sigma ^{\mu }\bar\theta ^{i}) +i (\theta _i \sigma ^{\mu }d\bar
\theta ^i),
\cr
&&\omega ^{\alpha }{}_i  = d\theta ^{\alpha }{}_i, \
\bar\omega _{\dot\alpha i} = d\bar\theta _{\dot\alpha i}.
\label{KartFormN2}
\end{eqnarray}
The use of the relation
\begin{equation}
{d\over d\tau } W =
\omega ^{\mu }_{\tau } \partial _{\mu }W +
\dot \theta ^{\alpha }{}_i D_{\alpha }{}^{i}W+
\dot{\bar \theta  }_{\dot\alpha i} \bar D^{\dot\alpha i}W
\label{FulDerW}
\end{equation}
allows to present $\hbox{S}_{\hbox{int}}$ (\ref{SN2int1}) in the
equivalent form
\begin{equation}
\hbox{S}_{\hbox{int}}=
i e \int\limits _{\Gamma ^*} d\tau \omega _\tau ^M A_M
+i \mu \int\limits _{\Gamma ^*} d\tau
\left( \lambda _2 \dot \theta ^{\alpha }{}_i D_{\alpha }{}^i W +
\bar\lambda _2 \dot {\bar\theta} _{\dot\alpha i} \bar D^{\dot\alpha i} \bar W
\right) .
\label{SLambda2}
\end{equation}
Following from Eq.(\ref{SLambda2}) is the existence of two physically
equivalent
ways for the representation of the required superfield $T_M$:
\begin{equation}
T_M = (\partial _{\mu }(\lambda _1 W + \bar \lambda _1 \bar W),\,0,\,0),
\label{TM1iii}
\end{equation}
and
\begin{equation}
T_M = (0,\, -i\lambda _2 D_{\alpha }{}^i W,\,
i \bar\lambda _2 \bar D^{\dot\alpha i} \bar W ).
\label{TM2iii}
\end{equation}
Without restricting generality of considerations,  we shall
use the $T_M$ representation in the form
\begin{equation}
T_M = (0,\,-{1\over 4} D_{\alpha }{}^i W,\,
-{1\over 4} \bar D^{\dot\alpha i} \bar W ).
\label{TMend}
\end{equation}
Note that the equivalent representation for $T_M$
(\ref{TMend}) is the following:
$T_M = ({1\over 4}\partial _{\mu }(W + \bar W),\,0,\,0)$.
Eq.(\ref{TMend}) creates the expressions
for the components of $N=2$ generalized superfield strength
of the Maxwell supermultiplet
\begin{eqnarray}
e_{(q)}{\Bbb F}^{(q)}{} _{\alpha }{}^i {}_{\beta }{}^j  &=&
e F_{\alpha }{}^i {}_{\beta }{}^j  =
-e \varepsilon _{\alpha \beta } \varepsilon ^{ij}\bar W,
\cr
e_{(q)}{\Bbb F}^{(q)}{} _{\dot\alpha i \dot\beta j}&=&
e F_{\dot\alpha i \dot\beta j}=
e \varepsilon _{\dot\alpha \dot\beta } \varepsilon _{ij} W ,
\cr
e_{(q)}{\Bbb F}^{(q)}{} _{\alpha}{}^i {}_{\dot\beta j}&=&
e F_{\alpha}{}^i {}_{\dot\beta j}
+i\mu D_{\alpha}{}^i T_{\dot\beta j}
+i\mu \bar D_{\dot\beta j} T_{\alpha }{}^i =
{1\over 2} \mu  \delta ^i_j \hat\partial _{\alpha \dot\beta }
(W+\bar W),
\cr
e_{(q)}{\Bbb F}^{(q)}{}_{\mu\alpha }{}^i &=&
e F_{\mu\alpha }{}^i+i\mu \partial _{\mu }T_{\alpha}{}^i
=e {i\over 4} (\sigma _{\mu }\bar D ^i )_{\alpha } \bar W
-{i\over 4}\mu \partial _{\mu } D_{\alpha }{}^i W ,
\cr
e_{(q)}{\Bbb F}^{(q)}{}_{\mu\dot\alpha i} &=&
e F_{\mu\dot\alpha i}+i\mu \partial _{\mu }T_{\dot\alpha i}
=-e {i\over 4} (D_i \sigma _{\mu })_{\dot\alpha } W-
{i\over 4}\mu \partial _{\mu } \bar D_{\dot\alpha i} \bar W ,
\cr
e_{(q)}{\Bbb F}^{(q)}{}_{\mu\nu}&=&e F_{\mu\nu}=
{e\over 16}(\bar D _i \tilde \sigma _{\mu\nu  }\bar D^i)\bar W +
{e\over 16}(D^i \sigma _{\mu\nu  } D_i) W .
\label{FwithLambda}
\end{eqnarray}
As a consequence of Eqs.(\ref{FwithLambda}), the algebra of $N=2$
doubly lengthened covariant derivatives $\nabla _M$ takes the form
\begin{eqnarray}
\{\nabla _{\alpha}{}^i, \nabla _{\beta}{}^j\}&=&
-e \varepsilon _{\alpha \beta } \varepsilon ^{ij}\bar W,
\cr
\{\bar\nabla _{\dot\alpha i}, \bar\nabla _{\dot\beta j}\}&=&
e \varepsilon _{\dot\alpha \dot\beta } \varepsilon _{ij} W ,
\cr
\{\nabla _{\alpha}{}^i , \bar\nabla _{\dot\beta j}\}&=&
2i \sigma ^\mu _{\alpha \dot\beta }\delta ^i_j \nabla _\mu
+{1\over 2} \mu  \delta ^i_j \hat\partial _{\alpha \dot\beta }
(W+\bar W),
\cr
[\nabla _{\mu } , \nabla _{\alpha }{}^i] &=&
e {i\over 4} (\sigma _{\mu }\bar D ^i )_{\alpha } \bar W
-{i\over 4}\mu \partial _{\mu } D_{\alpha }{}^i W ,
\cr
[\nabla _{\mu } , \bar\nabla _{\dot\alpha i}] &=&
=-e {i\over 4} (D_i \sigma _{\mu })_{\dot\alpha } W-
{i\over 4}\mu \partial _{\mu } \bar D_{\dot\alpha i} \bar W ,
\cr
[\nabla _{\mu } , \nabla _{\nu }] &=&
{e\over 16}(\bar D _i \tilde \sigma _{\mu\nu  }\bar D^i)\bar W +
{e\over 16}(D^i \sigma _{\mu\nu  } D_i) W .
\label{AlgDerNablaN2}
\end{eqnarray}

The motin equations of the charged $N=2$ superparticle with
AMM generated by the action (\ref{SLambda2}) and $S_0$
\begin{equation}
\hbox{S} = \hbox{S}_{0} + \hbox{S}_{\hbox{int}} =
{1\over 2}\int \limits_{\Gamma ^*} d\tau   \left[ {\omega _{\tau  }^{\mu }
\omega _{\tau \mu} \over  g_{\tau } }+
g_{\tau } m^2 \right] +
i \int \limits_{\Gamma ^*} d\tau  \, \omega _{\tau } ^M e^{(q) }
{\Bbb A}^{(q)}_M
\label{SfullN2}
\end{equation}
take the form of the generalized Lorentz equations
\begin{eqnarray}
(g_{\tau }^{-1} {\omega}_{\tau  \mu})\dot{\hphantom{)}}&=&
-i \omega _{\tau  } ^M e_{(q)}{\Bbb F}^{(q) }_{M\mu},\nonumber\\
g_{\tau }^{-1} {\omega}_{\tau  \mu} (\sigma ^{\mu}
{\dot{\bar \theta}}^i )_{\alpha}
&=&
{1\over 2} \omega _{\tau  } ^M e_{(q)} {\Bbb F}^{(q) }_{M{\alpha}}{}^i ,
\nonumber\\
g_{\tau }^{-1} {\omega}_{\tau  \mu} ({\dot\theta}_i
\sigma ^{\mu})_{\dot\alpha}&=&
-{1\over 2} \omega _{\tau  } ^M e_{(q)} {\Bbb F}^{(q) }_{M{\dot\alpha}i}
.\label{EqMoN2I}
\end{eqnarray}
where ${\Bbb F}^{(q)}_{M\mu }$ is defined by Eqs.(\ref{EqMoN2I}).

\section{}

The component expansion of the superfield $T _{\alpha }{}^i$
depending on the chiral variables has the form analogous
to \cite{West}
\begin{eqnarray}
T_{\alpha }{}^i (z_L)&=&
{i\over 4}\lambda _\alpha {}^i(x_L)+
{1\over 8} \theta _{\alpha k} C^{ki}(x_L)
-{1\over 8} \theta ^{\beta  i} f_{\beta \alpha }(x_L)
-{i\over 2} (\hat\partial \bar \theta ^i)_{\alpha } w (x_L)
-{1\over 2} (\hat\partial \bar\theta ^i)_{\alpha }
(\theta _j \lambda ^j (x_L))
\cr &&
- {1\over 4}\Xi ^{(2,0)ij} (\hat\partial \bar\lambda _j (x_L))_{\alpha }
- {1\over 4}\Xi ^{(2,0)\beta }{}_{\alpha }
(\hat\partial \bar\lambda ^i  (x_L))_{\beta }
+ {i\over 8}\Xi ^{(2,0)jk} (\hat\partial \bar\theta ^i)_{\alpha }
C_{jk}(x_L)
\cr &&
- {i\over 8}\Xi ^{(2,0)\beta\gamma} (\hat\partial \bar\theta ^i)_{\alpha }
f_{\beta \gamma }(x_L)
+{1\over 3} \theta _{\alpha }{}^i \Xi ^{(2,0)}{}_j{}^i \Box \bar w  (x_L)
\cr &&
+ {i\over 3}\Xi ^{(2,0)\gamma }{}_{\beta }
\theta ^{\beta }{}_j
(\hat\partial \bar\theta ^i)_{\alpha }
(\hat\partial \bar\lambda ^j  (x_L))_{\gamma }
+ {i\over 6}\Xi ^{(4,0)} (\hat\partial \bar\theta ^i)_{\alpha }
\Box \bar w  (x_L) .
\label{OmegaCompI}
\end{eqnarray}
The explicit form for the monomials $\Xi ^{(m,n)}$ is given
in \cite{ZTkpti1}.

The physical content of the theory is convenient to
analyse in the central base coordinates. In this base
the part of the action (\ref{SLambda2}) which describes
the contribution of AMM particle has the form
\begin{eqnarray}
\hbox{S}_{\hbox{int}}^{(\mu )}\Bigl| _{\hbox{\small photon}}&=&
- \mu \int\limits _{\Gamma ^*} d\tau \left( \omega _\tau ^M T_M
(x^{\mu },\theta ^{\alpha }{}_i ,\bar\theta _{\dot\alpha i})
\Bigl|_{\hbox{\small photon}}\right) =
\cr
&=&\mu \int\limits _{\Gamma ^*} d\tau
\Biggl[
i (\dot\theta \sigma ^{\mu \nu }\theta ) v_{\mu \nu }
+ {1\over 2} (\dot\theta _i \sigma ^{\nu }\bar\theta ^j)
\Xi ^{(2,0)i} {}_j \partial ^{\mu }v_{\mu \nu }
+{1\over 2} (\dot \theta _i \sigma ^{\mu \nu }\theta _k)
(\theta ^k \sigma ^{\rho }\bar\theta ^i)\partial _{\rho }v_{\mu \nu }
\cr
&&
-(\dot\theta \sigma ^{\rho }\bar \theta )(\theta \sigma ^{\mu \nu }\theta )
\partial _{\rho } v_{\mu \nu }
+{i\over 2} (\dot\theta _i\theta _j) \Xi ^{(0,2)ij}
(\theta \sigma ^{\mu \nu } \theta )\Box v_{\mu \nu }
+{i\over 2}(\dot\theta \sigma ^{\mu \nu }\theta )\Xi ^{(2,2)}
\Box v_{\mu \nu }
\cr
&&
+i (\dot\theta \sigma ^{\mu \nu }\theta )\Xi ^{(2,2)\rho \sigma }
\partial _{\rho }\partial _{\sigma }v_{\mu \nu }
-{i\over 2}(\dot\theta _i\sigma ^{\rho }\bar\theta _k)
(\theta ^i \sigma ^{\sigma }\bar \theta ^k)(\theta \sigma ^{\mu \nu }\theta )
\partial _{\rho }\partial _{\sigma }v_{\mu \nu }
\cr
&&
-{1\over 2} (\dot\theta \sigma ^{\rho }\bar\theta )
(\theta \sigma ^{\mu \nu }\theta ) \Xi ^{(2,2)}\partial _{\rho }
\Box v_{\mu \nu }
-(\dot\theta \sigma ^{\rho }\bar\theta )(\theta \sigma ^{\mu \nu }\theta )
\Xi ^{(2,2)\sigma \eta }\partial _\rho \partial _\sigma
\partial _\eta v_{\mu \nu }
\cr
&&
+{1\over 4}(\dot\theta _i \sigma ^\nu \bar\theta ^j )
\Xi ^{(2,2)} \Xi ^{(2,0)i}{}_j \Box \partial ^{\mu }v_{\mu \nu }
+{1\over 4}  (\dot\theta _i \sigma ^{\mu \nu }\theta _k)
(\theta ^k \sigma ^\rho \bar\theta ^i) \Xi ^{(2,2)}\Box \partial _{\rho }
v_{\mu \nu }
\cr
&&
+{1\over 2}(\dot\theta _i\sigma ^\nu \bar\theta ^j)
\Xi ^{(2,2)\rho \sigma } \Xi ^{(2,0)i}{}_j \partial _\rho
\partial _\sigma \partial ^\mu v_{\mu \nu }
+{1\over 2} (\dot\theta _i \sigma ^{\mu \nu }\theta _k)
(\theta ^k \sigma ^\rho  \bar\theta ^i)\Xi ^{(2,2)\sigma \eta }
\partial _\rho \partial _\sigma \partial _\eta v_{\mu \nu }
+ \hbox{c.c.} \Biggr].
\label{EqPhMagComp}
\end{eqnarray}

To elucidate the physical meaning of Eq.(\ref{OmegaCompI}),
note that it is a pseudoclassical limit $\hbar \to 0$
of field theory. Under such a consideration the
grassmannian spinors $\theta _{\alpha }{}^i$ are treated as a limit
of the fermionic Fock operators \cite{CasalbI,BerezII,SUSY_SGRA_10}
$b_{\alpha }{}^i$ which describe spin degrees of
freedom. Since $\theta _{\alpha }{}^i$ have dimensionality
$L^{-{1/2}}$ and $b_{\alpha }{}^i$ are choosen dimensionless, they
are connected by the relation
\begin{equation}
\sqrt{{\hbar \over 2mc}} \, b _{\alpha }{}^i
\, \to \, \theta _{\alpha }{}^i \ \hbox{when} \ \hbar \to 0 ,
\label{TransfQuan1}
\end{equation}
where $\hbar/{mc}$ is the Compton wavelength corresponding
to the massive Dirac field quants. Further it is convenient
to pass from $\theta _{\alpha }{}^i$ and
$\dot\theta _{\alpha }{}^i$ having the dimension $L^{1\over 2}$ to a
Dirac bispinor $\Theta ^i$
\begin{eqnarray}
\Theta  ^i = \left( {1\over g_{\tau }}\dot\theta _{\alpha } {}^i \atop
m^2 \bar\theta ^{\dot\alpha i}\right) ,
\qquad \Sigma _{\mu \nu } = {i\over 4} [\gamma _{\mu },\gamma _{\nu }],
\label{DefPsiSigmaN2}
\end{eqnarray}
where $\gamma _{\mu }$ are the Dirac matrices.
The bispinor $\Theta ^i$ has the dimension $L^{-{3\over 2}}$
and is an invariant under the proper time reparametrization
due to the presence of the einbein $g_{\tau}$ (\ref{SfullN2}),
which has the dimension $L^{-2}$ .
In the terms of $\Theta ^i$ (\ref{DefPsiSigmaN2})
the action (\ref{EqPhMagComp}) is presented as
\begin{eqnarray}
\hbox{S}_{\hbox{int}}^{(\mu )}\Bigl| _{\hbox{\small photon}}&=&
-\mu \int \limits _{\Gamma ^*}
\left( {g_{\tau } d\tau \over m^2}\right)    \left[
(\bar\Theta  _i \Sigma _{\mu \nu } \Theta  ^i )v^{\mu \nu }+
\hbox{high order rel. corrections}
\right] ,
\label{ActMuPhotonSigmaN2}
\end{eqnarray}
where $\left( {g d\tau \over m^2}\right)$ is a 1-dimensional
reparametrization invariant ``volume'' having the dimension $L^4$.
Due to the fact
that the grassmannian bispinor $\Theta ^i$ are treated as a
pseudoclassical limit of fermionic field operators $\Psi ^i$
\cite{CasalbI,BerezII,SUSY_SGRA_10},
the first term in (\ref{ActMuPhotonSigmaN2})
has a sense of a pseudoclassical limit of the Pauli term. The
latter describes the electromagnetic interaction of neutral
particles possessing the AMM equal $\mu $. This observation
explains the physical meaning of the constant $\mu $.

A possible field theory image of the action (\ref{ActMuPhotonSigmaN2})
can be restored by the substitution which conserves
the reparametrization symmetry
and all the dimensions
\begin{eqnarray}
\Theta ^i (\tau ) \, \to \, \Psi ^i (x),\quad
{g_{\tau } d\tau \over m^2} \, \to \,  d^4 x ,
\label{TransfQuant2}
\end{eqnarray}
where $\Psi ^i (x)$ has the sense of the field operator of a neutral
particle with spin $1/2$ possesing the canonical
dimensionality $L^{-{3\over 2}}$.

Passing to the photino part of the action 
and preserving the leading term of the expansion in powers
of $c^{-1}$, we find
\begin{eqnarray}
\hbox{S}_{\hbox{int}}^{(\mu )}\Bigl| _{\hbox{\small photino}}\!&=&\!
-i {\mu \over 4} \int\limits _{\Gamma ^*} g_{\tau } d\tau    \left[
\bar\Lambda _{Ri} \Theta _{L}{}^i - \bar\Theta _{Li} \Lambda _{R}{}^i
\right] \!\! ,
\label{ActMuPhotinoN2}
\end{eqnarray}
where $ \Lambda ^i{}_{ \left(  {L\atop R} \right) } =
{\displaystyle 1\pm i\gamma _5 \over \displaystyle 2} \Lambda ^i , \
\Lambda ^i = \left( {\displaystyle \lambda _{\alpha }{}^i \atop
\displaystyle \bar\lambda ^{\dot\alpha i}}\right)$ is the photino
field. The expression (\ref{ActMuPhotinoN2})
is a pseudoclassical image of the field action
\begin{eqnarray}
\hbox{S}_{\hbox{int}}^{(\mu )}\Bigl| _{\hbox{\small photino}}\!&=&
\!\!-i\, {\mu m^2 \over  4 } \int dx^4     \left[
\bar\Lambda _{Ri} \Psi _{L}{}^i - \bar\Psi _{Li} \Lambda _{R}{}^i
\right] \!\!,
\label{ActMuPhotinoN2Field1}
\end{eqnarray}
which can be interpreted as a hint for a possible conversion of the
massive neutralino $\Psi _L^i$ in to the photino $\Lambda _{Ri}$ with
the coupling constant proportional to the AMM of neutralino.

Note also that supersymmetry, together with the proposed extension
of the minimality principle, reproduces rather complicated, but
controlable structure of high-order terms in the expansion
with respect to $1/c$. This structure can be restored by the study
of the component structure of the extended superfield 
and the interpretation of $\Theta ^i$ as
pseudoclassical images of the spinor fields \cite{TZFR2}.

Thus, we conclude that the proposed here possibility of a
generalization of the minimality principle can be found helpful
for studying possible spin effects in the electromagnetic
interaction of charged and neutral fermions predicted
by supersymmetry \cite{SeibWitt}.

We are grateful to N.~Berkovitz, E.A.~Ivanov, A.P.~Rekalo,
J.P.~Stepanovskij and B.M.~Zupnik for fruitful discussions
and critical remarks.

This investigation was supported in part by INTAS Grant 93-127 and
93-633, INTAS and The Netherlands Government Grant 94-2317,
by the Foundation of the Ukrainian State Committee for Science and
Technology within the framework of the program for fundamental research
(No. 3/664), and  State Foundation for Fundamental Investigations
of Ukraine.


\begin{references}

\bibitem{Cartan1}
Cartan A.H.
Ann. Ec. Norm. Sup. {\bf 40}, 324 (1923); {\bf 41}, 1 (1924).

\bibitem{Volkov10}
Volkov D.V.
Phenomenological lagrangian of the Goldstone particles.
Preprint ITP-69-75. Kiev, 1969.

\bibitem{pionSUSY}
Volkov D.V., Akulov V.P.
Pis'ma Zh. Eksp. Teor. Fiz. {\bf 16}, 621 (1972).

\bibitem{SUSY_102}
Volkov D.V., Soroka V.A.
Pis'ma Zh. Eksp. Teor. Fiz. {\bf 18}, 529 (1973).


\bibitem{SUSY_101}
Volkov D.V., Zheltukhin A.A. and Tkach V.I.
Teor. Mat. Fiz. {\bf 10}, 329 (1972).

\bibitem{ShSch1}
Sherk J., Schwarz J.H.
Phys. Lett. {\bf B52}, 347 (1974).

\bibitem{Kalb}
Kalb M., Ramond P. Phys. Rev. {\bf D9}, 2273 (1974).

\bibitem{TZFR2}
Tugai V.V., Zheltukhin A.A.
Phys. Rev. {\bf D54}, 4160 (1996).

\bibitem{TZPL3}
Zheltukhin A.A., Tugai V.V.
Pis'ma Zh. Eksp. Teor. Fiz. {\bf 61}, 532 (1995).

\bibitem{WESS}
Wess J., Bagger J. Supersymmetry and Supergravity.
Prin. Univ. Press. Princeton. New Jersey, 1983.

\bibitem{West} West P.
Introduction to supersymmetry and supergravity.
World Scientific, 1986.


\bibitem{Lusanna}
Lusanna L., Milevski B.
Nucl. Phys. {\bf B247}, 396 (1984).

\bibitem{Sohnius}
Sohnius M., Stelle K., West P. Superspace and Supergravity.
Cambr. Univ. Press. Cambridge, 1980.

\bibitem{ZTkpti1}
Zheltukhin A.A., Tugai V.V.
Supersymmetry and minimality principle of electromagnetic interactions.
Preprint KPTI-97-6. Kharkov, 1997.

\bibitem{CasalbI}
Casalbuoni R. Nuov. Cim. {\bf 33A}, 389 (1976).

\bibitem{BerezII}
Berezin F.A., Marinov M.S.
Pis'ma Zh. Eksp. Teor. Fiz. {\bf 21}, 678 (1975).

\bibitem{SUSY_SGRA_10}
Akulov V.P., Volkov D.V.
Zh. Eksp. Teor. Fiz. {\bf 17}, 367 (1973).


\bibitem{SeibWitt}
Seiberg N., Witten E.
Nucl. Phys. {\bf B246}, 19 (1994).

\end{references}
\end{document}